# On the role of metaphor in information visualization

John S. Risch[1]

[1]*Pacific Northwest National Laboratory*
*Richland, Washington, U.S.A.*

**Correspondence:**
**John S. Risch**
**Future Point Systems, Inc.**
**1825 S Grant St**
**San Mateo, CA 94402, U.S.A.**
**E-mail: john@futurepointsystems.com**

**Abstract**

The concept of metaphor, in particular *graphical* (or *visual*) *metaphor,* is central to the field of information visualization. Information graphics and interactive information visualization systems employ a variety of metaphorical devices to make abstract, complex, voluminous, or otherwise difficult-to-comprehend information understandable in graphical terms. This paper explores the use of metaphor in information visualization, advancing the theory previously argued by Johnson, Lakoff, Tversky et al. that many information graphics are metaphorically understood in terms of cognitively entrenched spatial patterns known as *image schemas*. These patterns serve to structure and constrain abstract reasoning processes via metaphorical projection operations that are grounded in everyday perceptual experiences with phenomena such as containment, movement, and force dynamics. Building on previous research, I argue that information graphics promote comprehension of their target information through the use of graphical patterns that invoke these preexisting schematic structures. I further theorize that the degree of structural alignment of a particular graphic with one or more corresponding image schemas accounts for its perceived degree of intuitiveness. Accordingly, image schema theory can provide a powerful explanatory and predictive framework for visualization research. I review relevant theories of analogy and metaphor, and discuss the image schematic properties of several common types of information graphic. I conclude with the proposal that the inventory of image schemas culled from linguistic studies can serve as the basis for an inventory of design elements suitable for developing intuitive and effective new information visualization techniques.

**Keywords**: Graphical metaphor, analogy, image schemas, information design, perception, cognition

## Introduction

Information graphics[i] derive much of their power to inform and enlighten through the use of *graphical* (or *visual) metaphors*. For example, the length of a bar in a bar chart metaphorically represents a quantity of objects. Likewise, a tree diagram represents hierarchical relations existing among some set of abstract concepts. As with linguistic metaphors, visual metaphors map the characteristics of some well-understood source domain to a more poorly understood target domain so as to render aspects of the target understandable in terms of the source. In the case of information graphics, the source domain is a visuospatial pattern of some type, while the target domain is some characteristic of the information of interest. The question naturally arises, What are the characteristics of these spatial patterns that make them suitable as metaphorical source domains? In other words, what makes information graphics themselves meaningful, independent of phenomena they represent? Further, how does this meaning come about? Is it simply a matter of learned convention, or do we have some ingrained predisposition to understand certain graphical forms in particular ways? I argue here that the process of deriving meaning from information graphics likely involves both factors.

This paper advances the theory first argued by Johnson,[1] Lakoff,[2] and, more recently, Tversky,[3,4] that many information graphics are metaphorically understood in terms of *image schemas*, structural patterns that become established in our minds during early childhood in the course of our daily interactions with the world. According to image schema theory, early perceptual experiences with spatial phenomena such as containment, path-following, and object dynamics become generalized and subconsciously entrenched. These spatial frameworks are subsequently recruited for use in structuring and understanding more complex, abstract concepts. Image schemas are thus theorized to form a crucial link between perception and cognition, with obvious implications for information visualization.

To date, image schema theory and research has focused largely on the metaphorical use of spatial relations terminology in language. However, it seems likely that sign systems of all types have evolved to express conceptual relationships in image schematic terms, and, in turn, are interpreted according to their alignment with contextually appropriate schemas. This especially includes information graphics. In fact, the theoretically close structural and functional alignment of such graphics with these schemas may explain their uniquely powerful characteristics relative to other (e.g., linguistic) modes of expression.

While previous authors have largely limited their discussions to the image schematic characteristics of simple statistical graphics, I argue here that it seems likely that most, if not all, information graphics (in particular visualizations of abstract information) are ultimately understood in image schematic terms. I further argue that many interaction techniques commonly used in information visualization, such as progressive disclosure, can also be considered to have an image schematic basis. Consequently, image schema theory can provide a useful explanatory and predictive framework for guiding the design and evaluation of information visualization technologies.

In section 2 of this paper I review the prevailing theories of analogy and metaphor, including a discussion of the unique characteristics of graphical metaphors. In this section I make a crucial distinction between *analogical graphics*, which depict characteristics of inherently spatial phenomena, and *metaphorical graphics*, which depict abstract phenomena and are theoretically interpreted in image schematic terms. In section 3, I review image schema theory, including emerging evidence for their objective existence. This is followed, in section 4, by a discussion of the image schematic characteristics of a number of commonly used information graphics and interactive visualization techniques. In section 5, on the basis of the previous discussions, I propose that the standard inventory of image schemas derived from psychological and linguistic studies can serve as the basis for developing a graphical "grammar" suitable for information visualization purposes. I go on to discuss a proposed visualization design process that matches characteristics of the data and problem domains with graphical representations designed to invoke appropriate image schematic reasoning processes, and illustrate the concept with a hypothetical example. In section 6, I conclude the paper with a discussion of the implications of this approach, and of image schema theory in general, for future information visualization research.

## Analogy, metaphor, and graphical metaphors

It is now generally accepted that analogy and metaphor are key aspects of human cognition.[5,6,7] Previously considered aberrational and peripheral aspects of thought, analogical processes are now understood to play key roles in everyday communication, and underlie many, if not most, abstract reasoning processes.[8,9,10] We constantly use knowledge about things we already understand to help us make sense out of other, less well understood phenomena. Our capacity for metaphor, a special type of analogy, enables us to extend knowledge about things we understand to completely different domains of experience. *Graphical* metaphors enable us to understand such abstract concepts in terms of familiar and well-understood visuospatial phenomena.

Formally, analogy is defined as a cognitive process whereby the characteristics of a well-understood *source* (or *base*) *domain* are used to facilitate the comprehension of a more poorly understood *target domain*. This process involves a structure-mapping operation[ii] in which the source and target domain representations are aligned, and features of the source are projected to the target via inference.[11] According to Gentner's structure-mapping model of analogy comprehension, the features that are preferentially selected for projection are those that express systematic relations among relevant concepts in the source domain (ibid.).

In the structure-mapping model, domains and situations are represented as systems of conceptual objects, object-attributes, and inter-object relations. Knowledge is represented in the form of propositional networks of nodes and predicates in which the nodes represent objects (i.e., concepts), and predicates applied to the nodes express propositions about the associated concepts. Predicates that take a single object as an argument are referred to as *attributes* (e.g., YELLOW(x)), while those that take two or more arguments are referred to as *relations* (e.g., SMALLER THAN (x, y)). Gentner makes an additional key distinction between first-order predicates, which take objects as arguments, and second- and higher-order relations, which take propositions as arguments. For example, if HEAT(x, y) and WARM(y) are first-order predicates, CAUSE [HEAT(x, y), WARM(y)] is a second-order predicate. Structure-mapping operations can therefore involve not only individual objects, but also mappings between relations of objects and between relations of relations.

The analogy "A *T* is (like) an *S*" defines a mapping of characteristics of a source domain *S* to a target domain *T*. If the source domain is expressed in terms of nodes $s_1, s_{2,\ldots}\ s_i$ and predicates such as A, R, and R′, and the target domain has nodes $t_1, t_{2,\ldots}\ t_i$, then an analogy is defined as a mapping of selected object nodes of S onto the objects nodes of T:

$$M: s_i \dashrightarrow t_i \qquad (1)$$

Predicates from the source domain *S* are transferred to the target domain *T* where they are used to generate a candidate set of inferences about the nodes in *T*.

Gentner goes on to define a set of mapping rules that govern the transfer:

1. Ignore object attributes:

$$A(s_i)] \not\dashrightarrow [A(t_i) \qquad (2)$$

   Intuitively, this means that the specific properties of objects in the source domain are typically not used in making inferences about objects in the target domain. For example, in the analogy "a collection of hyperlinked documents is like a spider web," most of the physical properties of spider webs such as their stickiness, fragility, etc. are ignored in the transfer.

2. Try to preserve relations between objects:

$$R(s_i, s_j)] \dashrightarrow [R(t_i, t_j) \qquad (3)$$

   In the preceding example, the most salient characteristic of the source domain is the pattern of connectivity of the strands of the spider web, which is mapped to the pattern of connectivity among hyperlinked documents. In this case, the structural relationships among the component parts of spider webs are used to understand structural relationships among documents that reference one another.

3. In deciding which relations to preserve, preferentially select systems of relations rather than individual relations (the Systematicity Principle):

$$R'\,(R_1(s_i, s_j), R_2(s_k, s_l))] \dashrightarrow [R'\,(R_1(t_i, t_j), R_2(t_k, t_l)) \qquad (4)$$

   This is to say that we seek interpretations involving systems of interconnected knowledge over those involving independent facts. Source domains that contribute sets of relations that are themselves systematically related (i.e., that involve higher-order relations) support the generation of larger numbers of candidate inferences (and hence have more explanatory power) than those that contribute only individual relations. "An atom is like a tiny solar system" is a much "richer" analogy than "an atom is like a tiny ball."

There are several additional key features of the structure-mapping model. The first is that explanatory analogies normally involve a 1-1 mapping between nodes in the source domain and nodes in the target domain. The second is that relations in the source domain are understood to apply identically in the target domain. The third is that the structure-mapping model is purely syntactic in nature. The actual content of the relations is entirely variable, and may be static spatial information as in INSIDE(x,y), or constraint or dynamic causal information.

In analogical comparisons, the structure-mapping process favors rich, deep, contextually relevant systems of relational matches over shallow systems (the Systematicity Principle). Once an initial matching has been established, unmatched predicates present in the source structure but not in the target structure are considered as candidate inferences in the target. Analogy is therefore a form of induction. Because these candidate inferences serve to complete the target structure in a deep and systematically consistent way, they are often causally informative. As a consequence, analogical structure-mapping processes can lead to spontaneous and informative insights – the "ah-ha!" experience.[10] In this way, analogies enable new knowledge about the target domain to be acquired in a sometimes sudden and spectacular fashion.

Gentner distinguishes analogies from other types of domain comparisons according to the syntactic type of the shared versus nonshared predicates. Comparisons in which the ratio of mapped to nonmapped predicates are high, and which involve both object attribute and relational predicates, are referred to as *literal similarities*. An example of a literal similarity is "a mallet is like a hammer"; mallets and hammers share many material and structural characteristics. *Analogies* are comparisons involving relational predicates, but relatively few object attribute predicates, as previously discussed. Comparisons in which the converse is true, i.e., in which there is substantial overlap in object attributes but not in relations, are considered simply appearance matches. Finally, domain comparisons in which the source domain is an abstract relational structure, as in, for example, "a pulley is a simple machine," are referred to as *abstractions*.

Metaphor is a special type of analogy in which the source and target domains are semantically distant from one another.[7] In addition, concepts from the source and target domains are often blended together, rather than discretely mapped (ibid.). Metaphors are commonly signaled in language by the words "is a." For example, the phrase "an atom is like a tiny solar system" is an analogy; "all the world's a stage" is a metaphor. In both cases, however, knowledge about some better-understood entity or phenomenon is projected onto one that is less familiar or more complex or abstract in order to render it more intelligible.[iii]

Linguists differentiate among several subtypes of metaphor. Of most relevance to the present discussion is the concept of *dead* or *conceptual* metaphors.[2,8] These terms refer to metaphors that have become so entrenched in thinking and language that they go almost completely unnoticed in everyday use. For example, the phrase "grasp the concept" employs the conceptual metaphor IDEAS ARE OBJECTS (that can be grasped).[8] This metaphor is so established in our thought processes that its metaphorical nature passes unnoticed, as such, in actual use.[iv] As I will later argue, many of the graphical tropes employed in information visualization seem to be of this nature.

Metaphors and analogies are conceptual phenomena that can be expressed in multiple ways, for example linguistically (as in the previous examples) or graphically.[v] As with linguistic analogies, graphical analogies and metaphors make assertions about relations existing among some target set of concepts in terms of some source relational structure. In the graphical case, however, source relations are expressed in the form of a graphical construct of some kind. Graphics that qualify as *analogical* according to the structure-mapping model of analogy thus have the following characteristics:

1. They express systematic relations among the elements of a target domain in terms of those of some source domain.
2. They preserve only selected attributes of elements in the target domain.
3, There is a 1-1 mapping of elements in the source domain to elements in the target domain.
4. Relations in the source domain apply identically in the target domain.
5. The components of such graphics are syntactically neutral, that is, the same graphical elements can be applied to multiple target domains.

Graphical metaphors have the additional characteristics:

6. The target domain is semantically distant from the source domain.

7. Elements of the source and target domains may be conceptually blended, rather than discretely mapped.

Accordingly, information graphics which (spatially) depict inherently spatial phenomena, such as geographic maps, building blueprints, and molecular diagrams, can be considered *graphical analogues* of the things they depict. That is to say, such graphics preserve systematic spatial and topological relationships present among elements of the target domain, convey only selected attributes of those elements, and the graphics are semantically close to the target phenomenon. Such graphics serve to make the spatial phenomena they depict more understandable by filtering, compressing, rescaling, and otherwise abstracting salient aspects of the target domain in order to bring them within the scope of human perceptual and cognitive abilities.

In *graphical metaphors*, on the other hand, the source domain is a spatial pattern of some type that has clearly understood relational characteristics, but is semantically distant from the target phenomena that is depicted. This is to say that the spatial pattern includes discrete, perceptually salient components that bear some meaningful spatial or topological relationship to one another, but which are mapped to a (nominally) non-spatial target domain. As with graphical analogues, however, the relational characteristics of the spatial source domain are used to generate inferences in the target. In other words, graphical metaphors promote the understanding of abstract concepts in terms of well understood visuospatial phenomena of some kind.

One common type of metaphorical graphic maps knowledge about a source domain to some target domain by means of pictorial representations of physical objects and/or meaningful arrangements of such objects. The characteristics of the objects employed are transferred to the target domain where they promote inferences associated with the function of the object. Typically, both object attributes and relational predicates contribute, although (as with metaphors in general) relational predicates predominate. In addition, the objects employed are usually commonly encountered, human-scale artifacts chosen for maximum familiarity and hence maximal explanatory power. This type of graphic might be said to employ *mimetic* graphical metaphor in that its symbolic and relational characteristics stem from the characteristics of the physical objects that are depicted.

For example, a representation of a bridge might be employed in a business presentation to express the metaphor “this marketing plan is a bridge to the future.” The near and far ends of the bridge might be labeled “Now” and “Next Year,” respectively, while the bridge itself might be labeled “Marketing Plan.” In this way, the concept of a plan for achieving some desired result in the future is associated with the concept of a bridge connecting two physical locations that enables easy movement between them and/or the avoidance of some physical hazard. Physical movement between the two ends of the bridge is equated with a transition from one state to another. As a metaphorical source domain, the concept of “bridge” compactly encapsulates a nested system of relations involving notions of facilitated locomotion and hazard avoidance, as well as physical attributes such as strength and reliability. The target domain of “marketing plans” is clearly highly abstract and complex. In this case, the mapping promotes an understanding of only the most superficial aspects of the target domain in terms of equally superficial aspects of the source domain.

Graphical metaphors employing representations of physical objects have a long history of use, for example in art. Such depictions commonly involve the symbolic use of multiple objects arranged in meaningful spatial configurations that metaphorically (or allegorically) express relations among the concepts the objects represent. Renaissance painters, for example, used the positioning of symbolic objects in their compositions to express relations such as relative importance, superiority/inferiority and ordinality. These artists knew that certain spatial elements, patterns, and configurations would carry inherent meaning for their viewers, and constructed their illustrations accordingly. The situation with information graphics is analogous. Information designers routinely take for granted the fact that certain design elements and configurations have particular, abstract meanings. Even highly abstract graphics such as network diagrams depicting social relationships are immediately understandable. Why is it that certain spatial patterns are able to imbue ostensibly unrelated, abstract concepts with meaning?

The answer seems to be that we are conditioned from birth to organize and reason about abstract concepts in spatial terms. According to emerging theories of embodied cognition, early perceptual experiences with phenomena such as movement and force dynamics become abstracted and schematized. These schemas subsequently serve as "structural archetypes" for organizing and understanding more abstract and complex concepts. They are so cognitively entrenched as to be essentially "invisible" in practical use; in other words, they function as conceptual metaphors. Because they have a largely visuoperceptual basis, these structural patterns have been termed *image schemas*. According to theory, the metaphorical use of image schemas provides the basis for much of our conscious thought. The inherent spatial nature of many of our thought processes provides a strong rationale for the utility of information graphics, and, as I will argue later, may be responsible for the invention and gradual evolution of the variety of information graphics we see today.

## Image schema theory and the Spatialization of Form hypothesis

Image schema theory was first articulated by Johnson in a seminal text published in 1987,[1] and further elaborated by Lakoff that same year.[2] This work, in turn, emerged from Talmy[12,13,14] and Langacker's[15] empirical research on spatial-relations terms in language, as well as on Lakoff and Johnson's earlier theories regarding the importance of metaphor in human language and thought.[5,8] The essence of image schema theory is that recurrent patterns of kinesthetic and perceptual experience (e.g., physical movements, visual patterns) become subconsciously entrenched in our minds during early development. These patterns are subsequently recruited for use in abstract thinking. In other words, we use frequently encountered kinesthetic and perceptual patterns metaphorically to structure abstract concepts in thought and language. As I will argue, it seems likely that we use similar mechanisms in both constructing and interpreting graphical representations of abstract information.

In [16], Hampe summarizes image schema theory as follows:

- Image schemas are *directly meaningful* ("experiential"/"embodied"), *preconceptual* structures, which arise from, or are grounded in, human recurrent bodily movements through space, perceptual interactions, and ways of manipulating objects.
- Image schemas are highly *schematic* gestalts [that] capture the structural contours of sensory-motor experience, integrating information from multiple modalities.
- Image schemas exist as *continuous* and *analogue* patterns *beneath* conscious awareness, prior to and independently of other concepts.
- As gestalts, image schemas are both *internally structured*, i.e., made up of very few related parts, and highly *flexible*. This flexibility becomes manifest in the numerous transformations they undergo in various experiential contexts, all of which are closely related to perceptual (gestalt) principles.[vi]

According to theory, image schemas serve a special cognitive role in that they are both perceptual and conceptual in nature. They thus act as a bridge between our perceptual systems and "higher level" cognitive functions such as language and reasoning.[17] As Turner points out in [18], if we had evolved as amorphous, one-eyed creatures floating in liquid we would have no basis for forming concepts such as LEFT-RIGHT, UP-DOWN, NEAR-FAR, etc. But because we have evolved as bisymmetrical, binocular creatures in gravity, we naturally employ our bodily experiences as the basis for conceiving and describing more abstract concepts (e.g., s*he's in over her head, you're getting closer*).

Johnson characterizes image schemas as having a *gestalt structure* consisting of both highly schematic and relatively concrete and specific components. These two aspects can be considered analogous to the predicate-argument structures of propositional representations, with the maximally schematic components corresponding to arguments, and the more specific components corresponding to predicates. Deane notes that image schemas differ from propositional representations in a key way, however.[19] While traditional propositional representations signify only abstract truth conditions, image schemas have an experiential basis. As a consequence, they foster mental operations analogous to the actual physical operations on which they are based. Accordingly, they operate in congruence with their own unique experiential "logic."

Below are descriptions of several of the canonical image schemas defined by Johnson. Following Lakoff's format in [2], I include a description of the theoretical bodily basis for each schema, along with a description of the schema's structural components and several example metaphorical phrases illustrating their use in abstract concept structuring:

1. CONTAINMENT
   Bodily Experience: We experience our bodies as both containers for things (e.g., food) and things in containers (e.g., rooms).
   Structural Elements: An INTERIOR, a BOUNDARY, and an EXTERIOR
   Example Metaphors: *come into view*, *in a relationship*, *in debt*, *out of his mind*

2. PART-WHOLE
   Bodily Experience: We are whole beings with component parts we can manipulate.
   Structural Elements: A WHOLE with component PARTS, and a CONFIGURATION
   Example Metaphors: *part of the family, a piece of my mind, keep it together*

3. SOURCE-PATH-GOAL
   Bodily Experience: Physical movement involves starting and ending locations and a series of contiguous intermediate locations.
   Structural Elements: A SOURCE, a DESTINATION, a PATH, and a DIRECTION along the path
   Example Metaphors: *on the right path*, *don't get sidetracked*, *see where it leads*

4 LINK
   Bodily Experience: Eye "contact," umbilical cord, stream of consciousness
   Structural Elements: Two entities, A and B, with a connecting LINK
   Example Metaphors: *forming attachments, breaking ties, connecting the dots*

5. CENTER-PERIPHERY
   Bodily Experience: We experience our bodies and other objects as having centers and peripheries
   Structural Elements: An ENTITY, a CENTER, and a PERIPHERY
   Example Metaphors: *central idea, on edge, political moderate, fringe element*

6. BALANCE
   Bodily Experience: The experience of balancing during the act of walking
   Structural Elements: Two related entities, A and B, and a FULCRUM
   Example Metaphors: *weighing options, leaning towards it, balance of power*

It is important to understand that image schemas are not imagistic. That is to say, they are not mental representations of imagery. Rather, they are schematic representations of perceptual patterns, and therefore conceptual in nature.[1,20] Further, image schemas (often referred to as *kinesthetic image schemas*) are not founded on visual perception alone. Sound, touch, proprioception, and perhaps even smell may contribute to the establishment of schema useful for structuring abstract concepts. However, because the vision system is, by far, the most significant perceptual channel, image schemas are predominantly visually oriented. Finally, image schemas are also composeable, that is, base schemas can be combined to form more complex conceptual structures, and they can undergo mental transformations from type of schema to another.

In the years since Johnson first advanced this idea, there has been spirited debate in the research community over the defining characteristics of an image schema. While the debate continues, a general consensus has emerged regarding the standard inventory, shown in Table 1. The schemas listed below have been culled from a number of individual publications. A reference to the publication in which each schema was first described is provided for each entry. Note that Table 1 is not intended to be comprehensive or definitive, but only to provide a sense of the key concepts that have been discussed in the literature. In addition, I have taken some liberties in organizing the schemas in Table 1 into groups of related concepts that differ somewhat from previously published organizations.

The evidence for the psychological and neurological reality of image schemas is mounting, and is emerging from a wide variety of disciplines. Talmy's original linguistic studies in the 1970s and 80s established that spatial-relations terms in various languages could be analyzed in terms of combinations of a limited number of fundamental and universal schemas. Mandler's studies of learning in infants in the early 1990s found that image schema theory provided an excellent explanation for observed patterns of infant learning and conceptualization.[21] In 1995, Gibbs and Colston summarized empirical evidence from cognitive and developmental psychology studies that provided some of the first objective evidence for the "psychological reality" of image schemas.[22]

**Table 1.** Consolidated list of published image schemas. The reference numbers provided for each schema indicate the publication in which it was first described.

| | | |
|---|---|---|
| Existential | Spatial Motion | Balance |
| OBJECT[1] | ANIMATE MOTION[21] | AXIS BALANCE[1] |
| | CAUSED MOTION[21] | EQUILIBRIUM[1] |
| Spatial Relation | INANIMATE MOTION[21] | POINT BALANCE[1] |
| | LOCOMOTION[44] | TWIN-PAN BALANCE[1] |
| ABOVE[2] | SELF MOTION[21] | |
| ACROSS[2] | SOURCE-PATH-GOAL[1] | Process Dynamics |
| ADJACENCY[1] | | |
| CENTER-PERIPHERY[1] | Compositional | AGENCY[51] |
| CONTACT[1] | | CAUSE[51] |
| CONTAINMENT[1] | COLLECTION[1] | CYCLE[1] |
| COVERING[2] | COMPLEXITY[23] | CYCLIC CLIMAX[1] |
| (Relative) LENGTH[2] | FULL-EMPTY[1] | ENABLEMENT[1] |
| LINEAR ORDER[2] | LINK[1] | PROCESS[1] |
| NEAR-FAR[1] | MATCHING[1] | ITERATION[1] |
| (Relative) SCALE[1] | MASS COUNT[1] | |
| SUPPORT[1] | PART-WHOLE[1] | Force Dynamics |
| Spatial Form | Transformational | ATTRACTION[1] |
| | | BLOCKAGE[1] |
| COMPACTNESS[23] | EXPANSION[18] | COMPULSION[1] |
| PATH[1] | MERGING[1] | COUNTERFORCE[1] |
| STRAIGHT[49] | MULTIPLEX TO MASS[2] | DIVERSION[1] |
| SURFACE[1] | PATH FROM MOTION[2] | MOMENTUM[52] |
| ROUGH/BUMPY-SMOOTH[24] | PATH TO ENDPOINT[2] | RESISTANCE[52] |
| | PATH TO OBJECT MASS[2] | RESTRAINT[1] |
| Spatial Orientation | REFLEXIVE[2] | RESTRAINT REMOVAL[1] |
| | ROTATION[2] | |
| FRONT-BACK[2] | SPLITTING[1] | |
| LEFT-RIGHT[50] | SUPERIMPOSITION[1] | |
| UP-DOWN[2] | | |
| VERTICALITY[2] | | |

More recent research has shown that brain regions responsible for controlling motor actions are, in fact, actively engaged during abstract thought. For example, a recent series of psycholinguistic studies measured the time people took to process simple metaphorical phrases such as *push an issue*. These showed that people respond more quickly to metaphorical phrases that match immediately preceding actions (e.g., the act of physically grasping an object, followed by the phrase *grasp the concept*) than they do to phrases that don't match the action (e.g., kicking a ball, followed by the phrase *grasp the concept*).[23] This strongly suggests that the sensorimotor cortex is somehow involved in abstract thinking. In [24], Rohrer summarizes evidence from cognitive neurosciences research that shows that the sensorimotor cortical areas of the brain apparently do play a large role in semantic comprehension tasks. For example, functional magnetic resonance imaging (fMRI) studies show that it is possible to activate the somatomotor neural maps using not only perceptual, but also linguistic, input. Together, this evidence "support[s] the hypothesis that semantic understanding takes place via image schemata located in the same cortical areas which are known to map sensorimotor activity" (ibid.).

Lakoff's *Spatialization of Form hypothesis* is a general theory of conceptual organization that elaborates the role that image schemas play in semantic comprehension.[2] The Spatialization of Form hypothesis holds that conceptual structures used for organizing abstract concepts have an image-schematic basis.

Further, it associates particular key conceptual structures with specific image schemas, as follows:

- Categories (in general) are understood in terms of CONTAINMENT schemas.
- Hierarchical structure is understood in terms of PART-WHOLE schemas and UP-DOWN schemas.
- Relational structure is understood in terms of LINK schemas.
- Radial structure in categories is understood in terms of CENTER-PERIPHERY schemas.
- Foreground-background (i.e., salient vs. contextual) structure is understood in terms of FRONT-BACK schemas.
- Linear quantity scales are understood in terms of UP-DOWN schemas and LINEAR ORDER schemas.

Lakoff describes this organizing process as a metaphorical mapping of spatial structure into conceptual structure. In this way, concepts are given meaning according to the image schematic structures with which they are associated.

Image schemas play a role not only in semantic *comprehension* processes, but are also likely fundamental components of abstract *reasoning* processes, as well. According to this view, image schemas are employed in reasoning because they afford opportunities for mental operations such as pattern completion and mental simulation that are grounded in everyday experience. Recall that image schemas are generally propositional in form, characterized by predicate-argument structures that have associated inferential properties. Deane[19] illustrates this concept with a list of inferences commonly activated by the CONTAINMENT schema:

i) Containment protects contained objects from external forces;
ii) actions within a container are constrained by its boundaries;
iii) consequently, the location of a contained object is dictated by the location of the container;
iv) the container restricts one's ability to observe the contained object, depending on one's ability to see into the container;
v) containment is transitive: if A is in B, and B is in C, then A is in C.

Language that invokes the CONTAINMENT schema and instantiates it with entity-arguments therefore entails CONTAINMENT-specific inferences about those entities. Consequently, phrases such as "Joe is *in* the Army" entail inferences such as "Joe's actions are constrained" and "where the Army goes, Joe goes." Other schemas, for example SOURCE-PATH-GOAL, entail different inferences according to the physical experiences on which they are based. Graphics theoretically instantiate similar mental models. It seems likely, therefore, that much of the power of information graphics as reasoning tools stems from similar mechanisms.

## Image schematic characteristics of metaphorical information graphics

As previously discussed, information graphics that depict the characteristics of inherently spatial phenomena can be considered graphical analogs of the things they represent. However, many information graphics portray relations among non-spatial, abstract concepts in spatial terms, and are therefore metaphorical in nature. Many of these graphics seem to function by virtue of the fact that they employ graphical patterns that mimic spatial patterns that have become entrenched as image schemas. The forms and spatial configurations of lexical and graphical elements within information graphics foster the construal of mental models that structure the concepts associated with the elements according to familiar image-schematic patterns. In this way, the graphics become meaningful and are understood.

As a case in point, consider the schematic representations of bar charts shown in Fig. 1. The charts are presented unlabeled in order to emphasize their structural (i.e., syntactic, vs. semantic) characteristics. Fig. 1(a) shows a conventional example, depicting bars with aligned bases. This arrangement enables relative and absolute scale (i.e., length) readings to be performed via inspection of the unaligned "upper" ends of the bars. Fig. 1(b) shows bars of the same length, but with their tops aligned, rather than their bottoms. Subjectively, 1(b) is more difficult to interpret than 1(a), in terms of performing both absolute length readings and relative length comparisons. 1(b) also just "feels wrong." It can be said to be *counterintuitive*.

This seems due to the fact that vertically oriented bar charts, like many other common statistical graphics, are interpreted in terms of the metaphor MORE IS UP.[8] This metaphor is nearly universal across cultures and languages, reflected in English statements such as "the crime rate is going *up*" and "turn the heat *down.*" MORE IS UP consists of a mapping of a source domain that includes the UP-DOWN and SCALE image schemas into the target domain of relative quantities of things (not necessarily physical objects, but potentially also unitized abstract concepts such as *quarterly sales*). Experientially, we learn to correlate greater quantities with upward directionality in numerous ways. Adding objects to a pile or liquid to a container causes the surface to rise. As we mature physically (i.e., become "more") we increase in height (not *length*), as do other things in our experience, such as plants. According to image schema theory, these recurrent experiences become subconsciously entrenched, and subsequently serve to structure concept formation in analogous situations.

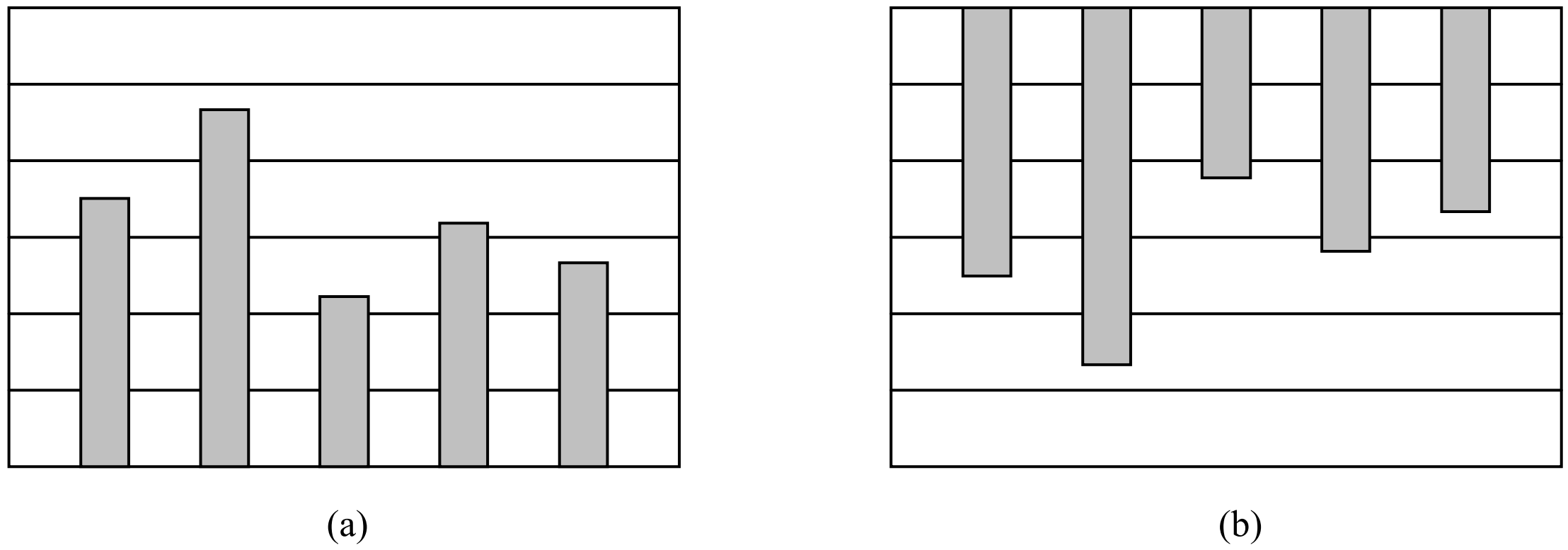

**Figure 1**. Comparison of otherwise identical bar charts employing MORE IS UP (a) vs. MORE IS DOWN (b) metaphorical mappings. Subjectively, (a) is quantitatively and comparatively easier to read, and feels more "intuitively correct," than (b).

The bar chart shown in Fig. 1(b) "feels wrong" because the alignment of the bar tops promotes interpretation in terms of the metaphor MORE IS DOWN, which is contrary to our everyday experience. In experience, vertically oriented items more commonly have their bases aligned than their tops, as we typically place them on some common surface (in the presence of gravity) for comparison purposes. Inverted bar charts such as that shown in Fig. 1(b) are occasionally encountered in practice, but are almost invariably used to indicate a decrease in some quantity (e.g., the price of a stock), and therefore still conform to the MORE IS UP metaphor. Horizontal orientations, of course, are also commonly employed for bar charts. However, unlike vertical charts, there is apparently no preferred (left-right) orientation.[3] This is likely due to the fact that, experientially, there is no perceived asymmetry to the world along the horizontal axis as there is in the vertical (ibid.).

Another example of a type of information graphic that has an image schematic basis is the Euler diagram. Several previous authors have noted that Venn/Euler diagrams employ a visual metaphor based on the CONTAINMENT schema.[2,4] In [17], Lakoff and Nunez argue that the embodied concepts of *collection* and *containment* underlie much of mathematics in general, and set theory, in particular. Recall that the structural elements of this schema include an INTERIOR, a BOUNDARY, and an EXTERIOR; this schema is depicted graphically in Fig. 2(a). As children, we learn the concepts of collection and containment early on through play, putting objects into containers and taking them out again, putting containers into other containers, etc.[20] As described in the previous section, these experiences ingrain in us a mental model of containment that we naturally employ in abstract reasoning. The novel invention of the idea of intersecting containers enabled the depiction of more complex logical relations such as those shown in Fig. 2(b). Note that we can conceive of, and reason about, the abstract concept of containment independent of the existence of any contained objects. This can be reflected in diagrams such as that shown in Fig. 2(c). Euler diagrams, like bar charts, are *intuitively understood* because they are directly construed in terms of embodied spatial schemata.

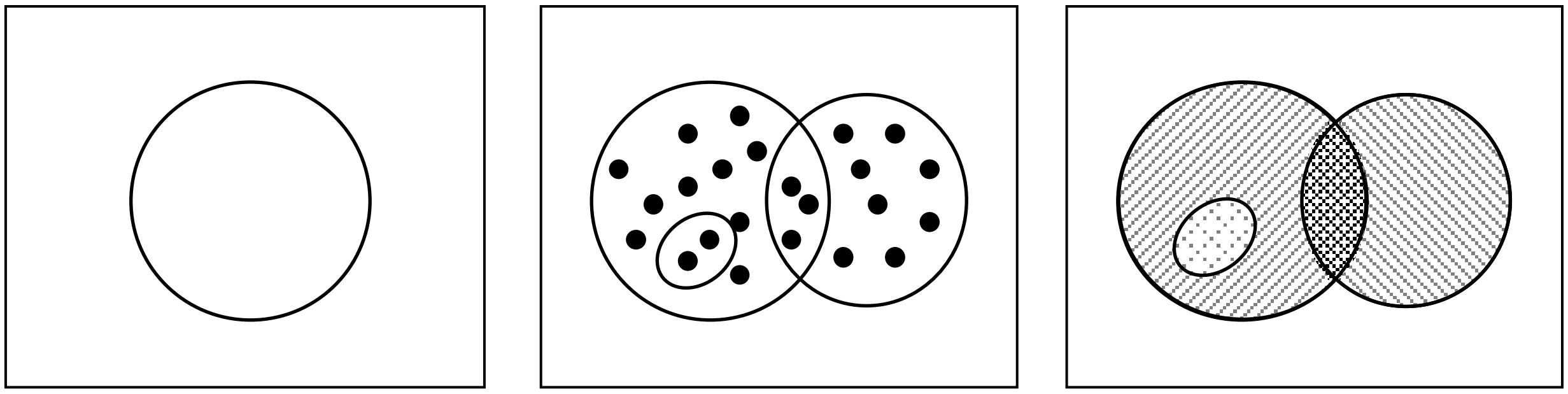

(a) (b) (c)

**Figure 2**. Venn/Euler diagrams illustrating interpretation of set membership by means of the CONTAINMENT schema-based visual metaphor. (a) schematic graphical representation of the CONTAINMENT image schema (b) OBJECT entities in nested and intersecting "containers," (c) a more abstract representation of intersection and containment

Network diagrams also seem to have an image schematic basis. However, network diagrams appear to have much deeper cognitive roots than the graphics discussed so far. According to Lakoff,[2] mental models (e.g., such as those instantiated in working memory by inspection of a graphic) have two key aspects, namely ontology and structure. Ontological elements may be either basic-level concepts such as entities, actions, states, etc., or they may be concepts characterized by cognitive models of other types (e.g., scripts). Structure in a mental model "consists of the properties of the elements and the relations obtaining among them" (ibid.). Recall that, according to the Spatialization of Form hypothesis, concepts are themselves conceived of in terms of the OBJECT schema, while semantic relations among concepts are understood in terms of the LINK schema. In other words, structural cognitive models[vii] have a graph-like structure defined in terms of the OBJECT/LINK schema pair. We think largely in terms of objects and relations, of "connections" among ideas.

The ubiquity of node/link diagram-based visualization tools testifies to the central role of the OBJECT/LINK schema in human thought. Examples include UML diagrams, flow charts, depictions of communications patterns, organization charts, "link charts," social network diagrams, and "mind maps," among a myriad of others. Although the semantics of these diagrams vary according to their ontological characteristics, their *structural* (i.e., schematic) aspects are similarly understood in terms of the general nature of the OBJECT/LINK image schema on which they are commonly based. These structural interpretations, in turn, theoretically guide interpretation of the ontological components of the diagrams to generate a gestalt interpretation of the diagrams as a whole.

The structure of a given network diagram can generally be interpreted in a number of different ways. Selection of a particular interpretive mode is likely driven not only by the ontological and structural characteristics of the design, but also by pragmatic factors such as viewer goals and context. Structure clearly plays a significant role in interpretation, however, and this structural interpretation likely has an experiential basis. Consider, for example, the series of simple schematic network diagrams depicted in Fig. 3(a-d). Fig. 3(a) shows an otherwise undifferentiated network defining a topological structure. Fig. 3(b) shows a PART-WHOLE interpretation, in which visual discontinuities such as curvature minima[25] (or possibly matching with geometric primitives[26]) guide interpretation of the network as an object with component parts. Figure 3(c) illustrates a SOURCE-PATH-GOAL interpretation in which meaning is derived in the form of a path traversal through the network.

As discussed in section 3, image schemas are closely associated with gestalt perceptual principles. It follows that node layout should significantly influence structural interpretation and attendant image schematic construal. Diagrams depicting identical network topologies but with differing node placements will result in different conceptualizations. For example, graph layouts such as those shown in Fig. 4(d) promote gestalt grouping of nodes into discrete sets according to spatial proximity, resulting in an entirely different PART-WHOLE interpretation than that shown in 4(b). Node layout thus has a profound effect on the ways in which the ontological characteristics of the associated nodes are construed.

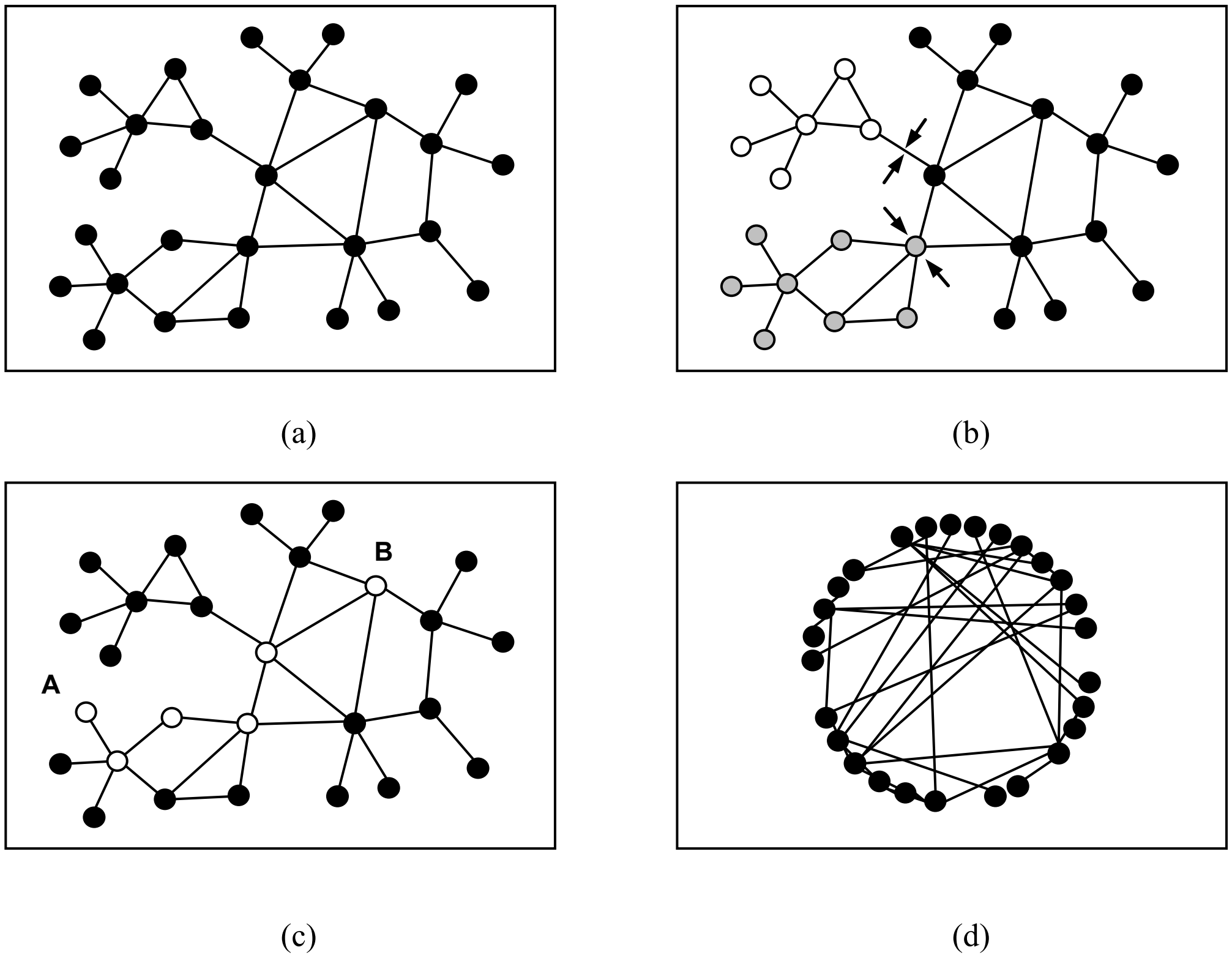

**Figure 3**. Examples of alternate image schematic interpretations of network diagram structure. (a) a simple node-link diagram, (b) illustration of a PART-WHOLE interpretation influenced by structural inflection points (arrows), (c) illustration of a SOURCE-PATH-GOAL interpretation involving path following from node "A" to node "B," (d) illustration of a node layout that promotes an interpretation according to gestalt perceptual grouping principles.

Hierarchical organization by generalization is another key strategy that humans use in conceptualization. Not surprisingly, developing strategies for depicting and interacting with hierarchically organized information has been a key focus of the information visualization community. According to the Spatialization of Form hypothesis, hierarchical structure is understood in terms of PART-WHOLE schemas and UP-DOWN schemas. PART-WHOLE relationship structure, in turn, is commonly (but not exclusively) understood in terms of OBJECT and LINK schemas. This OBJECT and LINK conceptual structure is frequently depicted graphically as a node/edge network structure. In the case of hierarchy diagrams, part-whole compositional structure originates from a single "root" node that conceptually represents the WHOLE.

Perhaps more interesting, however, is the fact that graphical representations of hierarchical conceptual structures commonly reflect a preferred vertical orientation that is in agreement with the Spatialization of Form hypothesis. In a survey of hierarchy diagrams in scientific textbooks, Tversky found that a vertical orientation was employed by 45 of 48 diagrams sampled.[4] Importantly, vertical relations among the entities depicted in the diagrams were significant, but horizontal relations were arbitrary. Further, the "root" of the hierarchy was almost invariably located at the top, and represented what Tversky referred to as a conceptual "ideal" of some kind. This preferred vertical and directed orientation suggests that some factor other than graphical convention is involved.

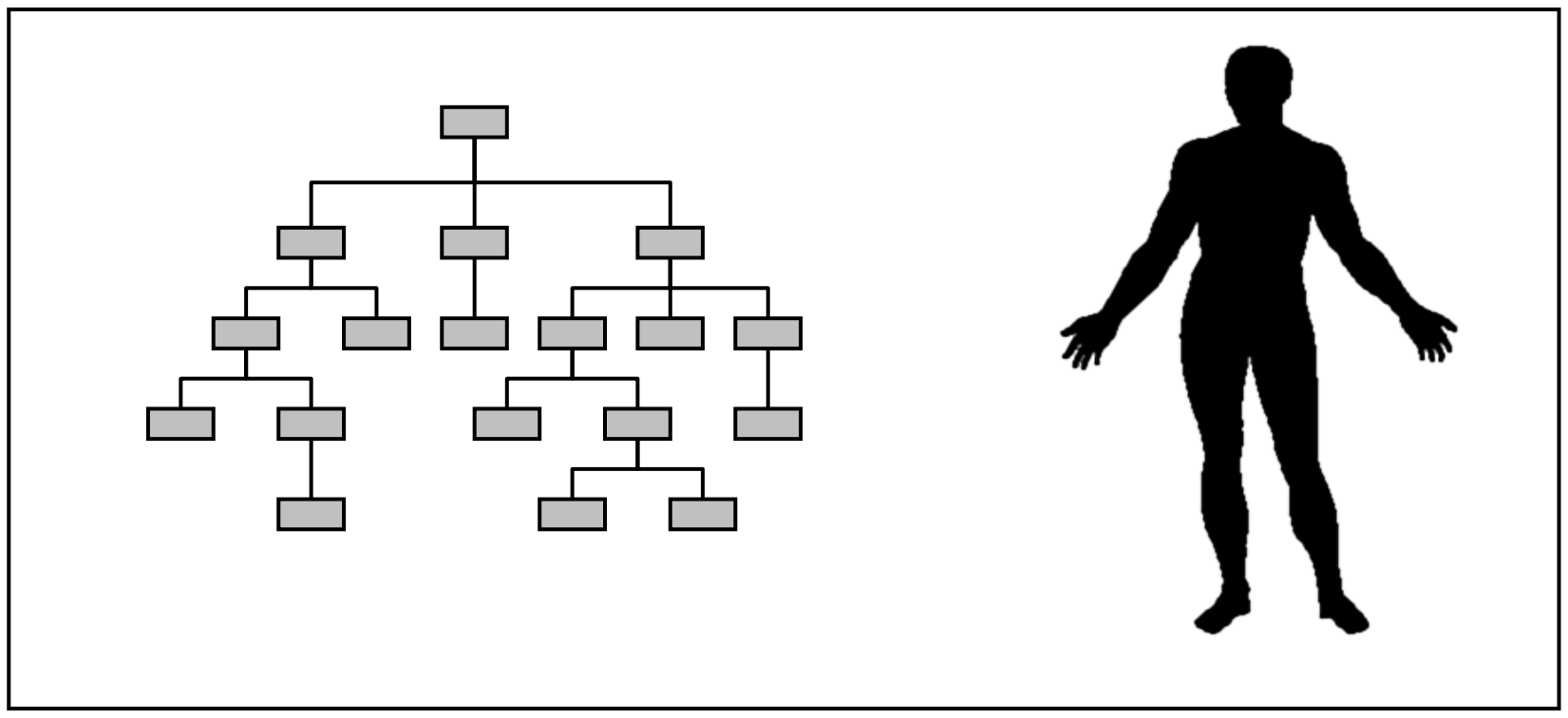

(a) (b)

**Figure 4**. A possible experiential source for the visual metaphor employed in vertically oriented hierarchy diagrams (a) is the human form (b).

It seems likely that such graphics have an image schematic basis. I speculate that the experiential source may be the human form (see Figs. 4a and 4b). The human figure possesses all the necessary criteria to serve as an image schematic source domain for this metaphor, including the facts that it is pervasively encountered in experience, is well understood, and is well and simply structured.[2] Further, the human form presents a hierarchical branching structure originating at the head, which is almost invariably located at the top of the body. Anecdotally, the head/face serves as the focus of attention during interpersonal interactions, and could thus arguably be considered the psychological "origin" of the human form, conceptually representing the WHOLE. Vertically oriented tree diagrams and the human form can thus be considered to be both structurally and semantically aligned.

A large variety of other graphical techniques have been developed for depicting hierarchically organized information. Many of these employ visual metaphors that also seem to have image schematic bases. For example, radial layouts such as those described in [27] and [28] theoretically promote interpretation of hierarchically organized PART-WHOLE structures in terms of the OBJECT/LINK and CENTER-PERIPHERY schemas, in which the "ideal" is located at the center of the display, with subordinate entities located at progressively greater distances away. Techniques such as the TreeMap[29] and related visualizations[30,31] seem to be based on the CONTAINMENT schema, depicting PART-WHOLE structure in the form of nested containers. Interestingly, image schema theory predicts that the substantially different bases for depicting PART-WHOLE relationships employed by OBJECT/ LINK vs. CONTAINMENT-based designs would cause them to be understood via substantially different processes. This seems to be born out in the results of user studies that show agreement of task proficiency along roughly schema-specific lines.[32]

So far, this discussion has centered on the image schematic characteristics of information graphics themselves. However, many of the interaction techniques that have evolved for use in information visualization software also seem to be founded on commonly encountered patterns of experience. For example, the technique of progressive disclosure in which additional detail is gradually revealed as a display is "zoomed" to a particular area of interest mimics the daily experience of perceiving greater detail in an object as we physically move closer to it. Conversely, as we physically move farther away from a collection of objects we eventually perceive them as fusing together into an indistinguishable visual mass, an effect reflected in the MULTIPLEX TO MASS schema. Graphical interactions with a mouse are analogous to physical object manipulations, and so forth.

Space prohibits a detailed analysis of the full range of information graphics and interactive visualization techniques in existence, but it seems likely that most other such methods ultimately have similar image schematic bases. This seems to be the case for two reasons. First, historically, information designers have naturally evolved graphic tools that mirror and reinforce their own thought processes. As a consequence, information graphics have come to mimic the subconscious schematic spatial patterns that form the basis of much of our thought and reasoning. Secondly, humans naturally seem to interpret spatial patterns, in particular purposefully designed ones, in terms of image schematic analogs. These two phenomena may have set up a mutually reinforcing condition that has resulted in the invention and refinement of the particular types of graphic devices we use today.

## General discussion

While definitions of the term *information visualization* vary,[33.34.35] there seems to be a consensus that the notion refers to the application of data visualization techniques to non-spatial or conceptual data. This is perhaps best captured in Card et al.'s definition of information visualization as "the use of computer-supported, interactive, visual representations of abstract data to amplify cognition".[33] The use of the phrase "visual representations of abstract data" serves to distinguish the practice of information visualization from the visualization of physical measurement or simulation data (e.g., *scientific visualization* or *cartography*). Further, the use of the term "amplify cognition" implies a goal of somehow augmenting abstract reasoning processes. If human thought processes are, in fact, largely metaphorical in nature, and if these metaphorical processes commonly employ image schemas as source domains, then image schema theory can theoretically serve as a basis for designing "cognition amplifying" visualization tools.

Fabrikant and Buttonfield[36] have previously suggested that graphical metaphors designed to invoke image schematic mental models can be used as an aid to information retrieval. I present here the stronger argument that the "experiential logic" associated with such metaphors can be leveraged to support not only search tasks, but also sensemaking. The key idea is that mental structures and processes that have evolved for *perceiving* and *doing* are routinely appropriated by the human mind for *understanding*, *knowing*,[37] and *reasoning*.[9] These structures and processes can, in theory, be purposefully activated via graphical metaphors that mimic appropriate image schematic patterns. Through judicious data transformations and careful alignment of problem domain, data, and metaphor semantics, information visualization tools might be developed that support reasoning about abstract concepts in ways that are similar to natural modes of reasoning about comparable perceptual phenomena. In a limited sense, information visualization tools based on such metaphors might be said to "augment" or "amplify" cognitive processes associated with abstract reasoning by fostering mental simulation and pattern completion operations appropriate to the problem domain and target information.

Metaphorical graphics ultimately seem to be founded on the base metaphor IDEAS (or MEANINGS) ARE OBJECTS.[8,38] The "graphical" version of this is IDEAS ARE GRAPHICAL OBJECTS. In addition, in keeping with the metaphor LINGUISTIC EXPRESSIONS ARE CONTAINERS (for ideas) (ibid.), I suggest that the metaphor INFORMATION GRAPHICS ARE CONTAINERS FOR IDEAS (represented as graphical objects) applies. The value of information graphics lies generally in their utility as forms of external memory and in their ability to support the generation of inferences associated with relations among ideas.

Concepts (IDEAS) in information graphics (both analogical and metaphorical) are commonly conveyed via conventional signifiers such as text labels or symbols (both in the sense of Peirce's definition of *symbol*),[39] or, as discussed earlier, via mimetic depictions of physical objects (Peirce's *icons* (ibid.)).[viii] While the meanings of iconic signs are (putatively) obvious, the meanings of nonlexical symbols (encoded, e.g., as shapes or colors) are typically conveyed through the use of a legend that maps (lexically expressed) concepts to corresponding signs. In an information visualization context, signifiers can represent concepts associated with the characteristics of individual information units (e.g., documents or database records), or with the aggregate properties of a collection of such units.

It is generally accepted that the principal function of information graphics is to depict relations among ideas or concepts. Bertin argues this quite forcefully when he states "What is properly called [information] graphics depicts only the relationships established among components or elements".[27] Tversky echoes this position, stating "…graphic elements are generally used to represent elements in the world, and graphic space is used to represent the relations between elements".[3] Information graphics therefore have two key aspects, namely 1) graphical objects representing concepts, and 2) a spatial configuration of those objects that expresses some systematic relation among their associated concepts. Consequently, I argue that information graphics fully qualify as either *analogues to* or *metaphors for* the information they represent according to Gentner's structure-mapping model of analogy. This is to say that not only do they 1) express systematic relations among the elements of a target domain in terms of those of a (spatial) source domain, but 2) also preserve only selected attributes of elements in the target domain, 3) employ 1-1 mappings of elements from source to target domain, 4) are syntactically neutral, and 5) relations in the source domain apply identically in the target domain.

Relations among concepts are expressed via meaningful spatial arrangements of lexical and nonlexical (i.e., GRAPHICAL OBJECT) concept signifiers. As with analogies in general, these signs and relations correspond to the arguments and predicates, respectively, of a propositional system. In the case of graphical metaphors, the predicates commonly correspond to experiential patterns associated with, e.g., the manipulation and behavior of physical objects. Similar to linguistic metaphors, new knowledge is derived in the form of inferences based on metaphorical projection of the characteristics of these patterns to the abstract phenomenon the graphics depict. Unlike linguistic metaphors, however, concepts and relations in metaphorical information graphics are likely processed by the independent "what" and "where" neuroanatomical pathways thought to be responsible for object identification and spatial relationship processing, respectively, in the human visuoperceptual system.[40,41] Theoretically, this fact may account for the subjective sense that information graphics enable a form of "parallel cognitive processing" of the concepts they present.

Meaning in metaphorical graphics is therefore considered a product of both the meanings associated with their component signifiers and that associated with the image schematic conceptual structures the graphics invoke. Collectively, these elements foster the construal of mental models in working memory that have image schematic structure and enable relationships among concepts to be understood via mental processes such as pattern completion and mental simulation. In the same way that, for example, "following" a "path" through a series of hyperlinked Web pages is construed as (egocentric) traversal of a physical path between contiguous physical locations,[ix] inspecting a network diagram depicting the hyperlink connectivity between those same pages enables an exocentric understanding of their structure as adjacent and connected locations in space. By presenting hyperlink structures as a metaphorical "map" that equates Web pages with physical locations, an information designer can draw upon the entrenched mental machinery used to understand physical orientation and movements through space to enable viewers to comprehend an otherwise abstract, complex and essentially incomprehensible phenomenon.

The inventory of image schemas presented in Table 1 can thus theoretically form the basis for a corresponding inventory of graphical metaphors, a kind of metaphorical "grammar" used to express conceptual relations among symbolically represented concepts. The information design process therefore becomes one of matching the characteristics of the target information to a graphical construct that fosters reasoning in terms of spatial metaphors and schemas that are aligned with the problem domain. This process notionally involves the following steps:

1) Identifying and/or deriving salient characteristics of the target information relative to the problem domain in question,
2) Characterizing the reasoning processes appropriate to both the target information and the problem domain,
3) Selecting one or more image schemas associated with the identified modes of reasoning,
4) Selecting an appropriate set of signifiers for representing the salient concepts of the target information, and
5) Generating a spatial representation that mimics the spatial structure of the selected image schema(s) and maps signifiers to concepts and spatial relations among signifiers to conceptual relations among concepts.

Theoretically, information graphics designed according to this process will promote construal of, and reasoning about, the chosen characteristics of the target information in ways that are both intuitive and appropriate to the natures of both the information and the analysis problem.

As an example, consider that abstract dynamic phenomena seem to be understood in image schematic terms. In [9], Lakoff summarizes the metaphorical nature of event structure as follows:

- States are bounded regions of space
- Changes are movements into or out of bounded regions
- Processes are movements
- Actions are self-propelled movements
- Causes are forces
- Purposes are destinations
- Means are paths to destinations

Problems involving dynamic phenomena, for example the conduct of business mergers and acquisitions, are thus commonly conceived of in terms of dynamic forces and spatial movements. An information graphic designed to support reasoning about the tradeoffs among optional courses of action might take advantage of this by appropriately mapping salient characteristics of the problem domain to graphical elements configured in ways that promote appropriate image schematic structuring and reasoning.

Such a graphic might, for instance, represent the initial and desired final states as bounded graphical regions of space. Alternative means of achieving the desired goal(s) might be represented as a series of parallel or intersecting paths connecting the starting and "destination" regions. Intermediate states associated with each course of action might be represented as sequential locations along the paths. The status of actors responsible for achieving the goals could be conveyed by symbols positioned at appropriate locations along the paths. The actual or hypothetical "progress" of the various actors toward or away from the goals could be portrayed as self-directed or caused movement (i.e., via animation). Impediments or enabling factors inhibiting or promoting the "progress" of the actors could be represented by graphical barriers or breaches, while arrows might be used to depict causal factors as forces impinging on the actors, or to represent causal relations among related events.[x]

Such a graphic would thus metaphorically represent concepts associated with a complex set of interlocking strategic options as the elements of a dynamic and systematically related graphical "landscape" that could be interpreted in terms of intuitive image schematic patterns such as SURFACE, CONTAINMENT, PATH TO ENDPOINT, SOURCE-PATH-GOAL, CAUSED MOTION, COUNTERFORCE, BLOCKAGE, RESTRAINT, RESTRAINT REMOVAL, etc. In this way, the information designer promotes reasoning operations appropriate to the target information through the use of spatial patterns and animated movements that are structurally and semantically aligned with entrenched modes of reasoning about generic events.

As a final thought, consider that it is generally taken for granted that information graphics should be *intuitive*. That is to say, they should graphically depict salient aspects of information in ways that enable them to "[attain] to direct knowledge or cognition without evident rational thought and inference."[xi] In fact, much of the supposed utility of information visualization technologies can be said to stem from their purported ability to make abstract and complex phenomena intuitively understandable. Image schema theory provides a basis for understanding and characterizing the nature of this intuitiveness. Theoretically, graphics that clearly and concisely invoke image schematic structuring of the concepts they depict will be perceived as more "intuitively meaningful" than those which do not. If the metaphor itself is graphically intuitive and has been appropriately mapped to the information, then, in theory, the target information itself will also be understood "without evident rational thought."

Image schema theory provides a theoretical explanation for how and why information graphics become meaningful to a viewer, as well as an explanation for why certain graphics seem more intuitive than others. As presented here, it also supports a number of falsifiable predictions with regard to information graphics, including the following:

1) Graphics that mimic image schematic patterns but that lack concept-signifying elements (i.e., that are purely schematic, e.g., Figs. 1-4) will still be construed as meaningful by viewers.
2) Graphics that are image-schematically aligned will be judged subjectively as being more "intuitive" than those that are not.
3) Image schema-aligned graphics will be learned more rapidly and with less effort than those that are misaligned or non-aligned.
4) Graphics that align interpretation requirements with image-schemas having correspondingly

appropriate experiential "logic" will prove to be more effective reasoning tools than those that do not.

Systematic psychological and software user studies are needed to validate or disprove these predictions before the full usefulness of the approach outlined here will be understood.

While the application of image schema theory to the visualization of abstract information appears to be a potentially fruitful avenue of research, the ultimate value of this approach remains unclear. Arbitrary graphical patterns and interactions can become meaningful via learning. While it seems unlikely, it may be that the "intuitive" nature of image schema-aligned graphics only confers an advantage (if any) during learning, and that, with use, arbitrary visualization techniques become equally effective. Along these lines, Cooper[42] has mounted a famously spirited attack on the use of high-level metaphors (such as the "desktop paradigm") in user interface design, arguing that learned, "idiomatic" paradigms are preferable. The utility of high-level metaphors as the basis for general human-computer interactions remains debatable. However, the user interface components Cooper describes as paragons of idiomatic design (e.g., scroll bars, splitters, combo boxes) are all, in fact, based on experiential patterns of interaction with physical objects, and are generally metaphorical in nature. Metaphors such as these are so deeply ingrained in human thought processes that they are essentially inseparable from the concepts with which they are associated, a fact that ultimately accounts for much of their utility. Finally, image schemas are only one type of schema used in visuospatial reasoning.[43,44] Aspectual schemas, scripts, etc. are all complementary or alternate candidates to serve as the basis for computerized tools for "amplifying cognition."

## Summary and conclusion

To say "*that* is like *this*" is to attempt to make a poorly understood phenomenon more comprehensible by relating it to a more meaningful one. This is exactly what information graphics do. Such graphics present selected aspects of complex phenomena in ways that are more comprehensible than their original form. This is accomplished via structure-mapping processes in which discrete elements of a graphic map to selected natural or derived characteristics of the target phenomena in ways that equate spatial relations in the source graphic to spatial or non-spatial relations in the target. With repeated use, these mapping processes often become so ingrained as to function instead as a type of categorization. When this occurs, the analogical or metaphorical nature of the process becomes effectively invisible, and "the picture becomes the thing" it represents.

I have argued that a primary distinction can be made between *analogical* graphics and *metaphorical* graphics. The two types differ according to the nature of the phenomena they represent. Analogical graphics depict inherently spatial phenomena in ways that preserve natural spatial relations present in the target. Such graphics serve to bring these phenomena within the limits of human perceptual, memory, and cognitive abilities. Thus geographic maps filter, abstract, and compress the spatial relationships among features of real terrains into human-scale artifacts that can be directly comprehended. Similarly, diagrams of microbial cell or molecular structures make selected spatial characteristics of the phenomena they depict perceivable and thus understandable. Such graphics are considered analogical because the source and target domains are both spatial in nature, and the spatial relations among elements are identical in source and target.

Metaphorical graphics, on the other hand, present non-spatial concepts in spatial terms. These relate abstract concepts in systematic ways that stimulate natural modes of conceptual structuring. Building on previous theory, I have argued that metaphorical graphics derive much of their meaning from their functional alignment with *image schemas*, cognitively entrenched patterns of physical experience that theoretically serve as a key bridge between perception and cognition. These patterns serve to structure and constrain abstract reasoning processes via metaphorical projection operations that are grounded in everyday cognitive and perceptual experiences. I have suggested that metaphorical graphics promote the image schematic structuring of concepts in working memory by mimicking the visuospatial patterns responsible for the schema's original formation. The sudden insights information graphics often induce – the "a-ha!" experience – can be theoretically explained as the inferential completion of such image schematic patterns.

Consequently, I have proposed that the standard inventory of image schemas derived from linguistic and cognitive studies can serve as the basis for developing a kind of visualization "grammar." Such a grammar would employ graphical analogs of image schematic patterns as syntactic elements that relate concepts expressed using conventional signifiers such as text, color, and symbology. Theoretically, graphics that purposefully align data and problem characteristics with appropriate image schematic graphical patterns can stimulate pattern completion and/or mental simulation processes that generate valuable new insights into the target phenomena they depict. I have presented a number of falsifiable predictions that can be used to test the validity of the theories presented here.

The use of the words "on" and "in" in the title of this paper is testament to both the ubiquity and invisibility of spatial metaphors in everyday thought and language. If information visualization tools truly hold the potential to amplify cognition, such tools will employ graphical devices that in some way augment everyday modes of thought and reasoning. The centrality of spatial metaphor in human cognition suggests that image schema theory can provide a useful theoretical framework suitable for guiding the design and evaluation of such visualization tools.

## Endnotes

[i] In the following discussion I will use the phrase "information graphics" to refer to static graphical representations of information and the phrase "information visualizations" to refer to the computer-supported dynamic, interactive variants of information graphics, after Card et al.[33].

[ii] The High-level Perception (HLP) model of Chalmers et al.[45] has been portrayed as a competitor to the structure-mapping model. However, Morrison and Dietrich[46] point out that the two models, in fact, address different phenomena. They characterize HLP as a theory of analogy production, whereas structure-mapping is a theory of analogy comprehension. As the focus of the present discussion is the comprehension of graphical metaphors, structure-mapping theory is the most appropriate model. It is clear, however, that perceptual processes are central to both the production and comprehension of graphical metaphors.

[iii] Poetic metaphors such as "the moon is a ghostly galleon" are highly complex phenomena that, while consistent with the structure-mapping model, are beyond the scope of the present discussion. For a comprehensive analysis of cognitive phenomena associated with poetic metaphors see [47].

[iv] In [48], Bowdle and Gentner propose a mechanism by which conceptual metaphors arise. They theorize that novel metaphors are initially interpreted via structure-mapping processes, but gradually become entrenched as general schemas with repeated exposure. Once entrenched such metaphors are conceptualized as a kind of category.

[v] Theoretically, metaphors can be expressed via any sensory modality. Sonification techniques, for example, can be considered as metaphorical expressions if they express data patterns in terms of auditory patterns that have previously understood relational characteristics.

[vi] All italics in original.

[vii] As opposed to functional or other types of mental models.

[viii] Peirce's *indexical signs* are also frequently employed in information graphics, for example, in the form of text labels associated with graphical elements by their adjacency.

[ix] The metaphor of physical movement employed in Web browsing is so strong it even extends to the possibility of "getting lost."

[x] See Tversky's comments regarding the use of arrows in information graphics to portray causal relations.[3]

[xi] Definition of "intuition," Webster's Third New International Dictionary.